# Morpho-anatomical characterization of the urogenital schistosmiasis vector Bulinus truncatus (Audouin, 1827) (Heterobranchia : Bulinidae) from Southwestern Europe

Alberto Martínez–Ortí[1,2], Sonia adam[1], Giovanni Garippa[3], Jérôme Boissier[4], M. Dolores Bargues[1,5] & Santiago Mas–Coma[1,5]

[1]Unit of Sanitary Parasitology, Department of Parasitology, Faculty of Pharmacy, University of Valencia, Av. Vicent Andrés Estellés s/n, 46100 Burjassot, Valencia, Spain

[2]Museu Valencià d'Història Natural–i\Biotaxa, Alginet, Valencia, Spain

[3]Unità di Malattie Parassitarie, Dipartimento di Medicina Veterinaria, Università di Sassari, via Vienna, 2, 07100 Sassari, Italy

[4]IHPE, Univ. Montpellier, CNRS, IFREMER, Univ. Perpignan Via Domitia, Perpignan, France

[5]CIBER de Enfermedades Infecciosas, Instituto de Salud Carlos IIII, C/Monforte de Lemos 3-5, Pabellón 11. Planta 0, 28029 Madrid, Spain

**Abstract**

Urogenital schistosomiasis has been present naturally in the South of Europe since the beginning of the 20th century and nowadays its presence is also known, at least imported by Sub-Saharan emigrants and tourists, in France, Italy, Portugal and Spain. One of the intermediate hosts of this trematode present in Europe is the bulinid mollusc *Bulinus truncatus*, non–native species that can be reached to Europe by humans and birds. In order to know this mollusc better, we carried out a morpho–anatomical study, of the shell, the reproductive system, radula, the respiratory organs and pseudobranch of several populations from Italy, France and Spain. Spanish conchological material studied comes from different populations, from material deposited in the "Museo Nacional de Ciencias Naturales" of Madrid and the "Museu de Ciències Naturals" of Barcelona, as well as from its own material deposited in the "Museu Valencià d'Història Natural" of Alginet (Valencia). The shell growth in captivity and the estimation of the population age of *B. truncatus* from El Ejido (Almería, Spain), has also been studied. Finally, the finding of aphallic and euphallic specimens in the different populations of southern Europe studied is presented and taxonomic and ecological data of the genus *Bulinus* are shown.

**Introduction**

It is evident that the world has been facing a new scenario of emerging infectious diseases in the last two decades (Jones *et al*., 2008), including many threats to human health and potential repercussions on global stability (Morens & Fauci, 2013). Parasitic diseases are included within the so–called Neglected Tropical Diseases (NTDs) by the World Health Organization (WHO, 2010). One of them is human schistosomiasis caused by infection with the trematode *Schistosoma* Weinland, 1858. It is estimated that at least 240 million people worldwide are infected, 90% of them in Sub–Saharan Africa, and that it causes around 300,000 deaths annually (WHO, 2010, 2013). Two species, *Schistosoma haematobium* (Bilharz, 1852) and *Schistosoma mansoni* Sambon, 1907, are responsible for urogenital and hepatointestinal schistosomiasis respectively in humans (Malek & Cheng, 1974; Brown, 1994).

Urogenital schistosomiasis appears limited to Africa and a part of the Near East (Doumenge *et al*., 1987). Unexpectedly, however, an outbreak of urogenital schistosomiasis has recently been detected on the French Mediterranean island of Corsica (Berry *et al*., 2014; Holtfreter *et al*., 2014). An overlap of climate change and global change factors appear to underlie this disease introduction (Boissier *et al*., 2015; Kincaid–Smith *et al*., 2017). Three schistosomes have been proved to be introduced into Corsica island, namely *S. haematobium*, *S. bovis* (Sonsino, 1876) and *S. haematobium–S. bovis* hybrids. Studies predict that these schistosome introductions to the South of Europe occur by Sub-S aharan emigrants and tourists in France, Italy, Portugal and Spain (Boissier *et al*. 2015; Salas–Coronas *et al*., 2018, 2019; Martínez–Ortí *et al*., 2019). The molecular analyses of these imported schistosomes indicated that the introduction most probably occurred from the northern part of Senegal (Boissier *et al*., 2016). Recently, Salas–Coronas *et al*. (2021) indicate the autochthonous transmission of urogenital Schistosomiasis in the province of Almería (Spain). Snail vectors are useful markers for the study of the impact of both climate change and global change on the transmission, epidemiology and also the spreading power of these trematodiases, mainly by increasing the cercarial production (Mas–Coma *et al*., 2008, 2009; Kincaid–Smith *et al*., 2017). The freshwater snail vectors of schistosomiasis are of the genus *Bulinus* O.F. Müller, 1781, and *Bulinus truncatus* (Audouin, 1827) vectorise urogenital schistosomiasis in Europe (Berry *et al*., 2014; Holtfreter *et al*., 2014; Boissier *et al*., 2015; Martínez–Ortí *et al.,* 2015, 2019). The underlying high and fast multiplication of these species is in great part due to their selfing capacity, mainly in aphallic specimens, found regularly in *B. truncatus* (Brown, 1994; Dillon, 1994). Schistosome– snail specificity is a complex phenomenon which involves the geographical location and macromolecular glycoproteins produced by the snail through its mucous secretions that provide a mechanism for host recognition by miracidia as in *Bulinus globosus* (Morelet, 1866) (Haberl *et al*., 1995; Allan *et al*., 2009). Consequently, the studies for the aforementioned purposes should not only focus on the bulinid vector but also involve an in–depth analysis of their environmental characteristics in the area or region where the gastropods are emerging or spreading. In fact, *B. truncatus* has a wide geographic range, extending along the coasts of the circum–Mediterranean region, a large part of Africa, the Middle East and the Turanic region. In Europe it has been cited from various countries such as Spain, mainland France and Corsica, mainland Greece and Crete, Italy (Sardinia and Sicily), Malta and Portugal (Brumpt, 1929; Germain, 1931; Larambergue, 1939; Coluzzi *et al*., 1965; Deiana *et al*., 1966a, 1966b; Berry *et al*., 2014; Holtfreter *et al*., 2014; Martínez–Ortí *et al.,* 2015, 2019).

The parasitological and veterinary interest of *B. truncatus* does not only concern its capacity to transmit *S. haematobium* and *S. haematobium– S. bovis* hybrid to humans and *S. bovis* to cattle, since it can transmit other species of trematodes such as *Paramphistonum cervi* (Schrank, 1790), *P. microbothrium* Fischoeder, 1901 or *Echinostoma* spp. (Martínez–Ortí *et al*., 2019). The susceptibility of the European *B. truncatus* to African schistosomes opens the query about the risk of introduction of the African *S. haematobium* into southern countries of Europe where this bulinid species is also known to be present and countries where African immigration is a continuous movement (Martínez–Ortí *et al*., 2015, 2019; Boissier *et al*., 2016).

Following with the risk map of urogenital schistosomiasis made by Martínez–Ortí *et al*. (2019) for the Iberian Peninsula and Balearic Islands, the present study has the aim to contribute to the knowledge of different aspects of *B. truncatus* populations from the Western Mediterranean frame: El Ejido in Almería province (Spain) and Villena in Alicante province (Spain) (Martínez– Ortí *et al*., 2015), Porto–Vecchio in Corsica (France) and San Teodoro in Sardinia (Italy), including comparisons with samples deposited in several Spanish museums such as the "Museo Nacional de Ciencias Naturales" of Madrid (MNCN) and the "Museu de Ciències Naturals" of Barcelona (MCNB), as well as samples deposited in the "Museu Valencià d'Història Natural" of Alginet (MVHN).

Aspects studied in the current paper are among those of interest for snail vector specimen identification, geographical strain differentiation, population age estimation, epidemiological characterization and disease transmission follow–up studies, and include (i) the morphology and measurements of the shell, (ii) the shell growth followed experimentally in laboratory–borne and reared specimens, (iii) anatomical features comprising the respiratory system and digestive organs such as the pseudobranch, jaw and the radula, (iv) ecological characteristics of the environment of the populations and (v) to improve the capacity to identify and monitor populations of *B. truncatus* in South of Europe and the potential introduction and spreading of *S. haematobium* and *S. haematobium–S. bovis* hybrid.

**Materials and Methods**
*Material collected*
The specimens from Almería (Spain) were collected on December 5 2014 and July 8 2015 in a pond that occupies a large artificial excavation in the course of a ravine from the Sierra de Gádor (El Ejido, Almería). The specimens from Corsica were collected on July 15 2015 and those from Sardinia on July 11 2015. All the material was preserved in 70% ethanol for morph–anatomical studies and 96% ethanol for molecular studies and are deposited in the MVHN of Alginet (Spain): El Ejido (Almería, Spain) (UTM 30SXH8572) (MVHN–071214TP04), Cavu river in Porto–Vecchio (Corsica, France) (UTM 32TNM2717) (MVHN–240815AF01), and Stagno Brandinchi of San Teodoro (Sardinia, Italy) (UTM 32TNL5720) (MVHN–031215MA06). The details of their habitats appear in Martínez– Ortí *et al*. (2015, 2019). The water data was collected using Hanna HI98312 instrument for temperature and conductivity, the Hanna HI8424 for pH and the Hanna HI9147 for dissolved oxygen meter.

*Material from museums*
Shells of several samples of *B. truncatus* deposited in MNCN of Madrid have also been studied and measured and the MCN of Barcelona also photographed. The four samples of the MNCN from the Ortiz de Zárate collection and Cobos collection are: "Barcelona, en una finca particular" (MNCN–15.05/40640) with five specimens recollected by Carlos Altimira determined as *Bulinus contortus* and "Barcelona, Huerto en la Bonanova" (MNCN- 15.05/26320) with three specimens collected in February of 1957 by Carlos Altimira and determined as *Isidora contortus*; the sample MNCN–15.05/26314 from Menorca island with two small specimens from Cobos collection and the sample MNCN– 15.05/40630 from "Palma de Mallorca" with 19 specimens. Of the samples from Barcelona (MNCN) only the four bigger specimens have been measured (Table 1). Another sample for conchological studies from MVHN corresponds to "Laguna de Villena" (Alicante, Spain) (MVHN- 281214TY01) (Fig. 7). The pseudobranch studied of *B. truncatus* comes from El Ejido while the pseudobranch of *Helisoma duryi* (Wetherby, 1879) from the Ágora of Valencia (Spain).

*Morpho–anatomical characters*

All obtained measurements of the shells come from literature and our measurements: length, (L), width, height of aperture (HA) and HA/L relation (Table 1). The ornamentation of the protoconch and the teleoconch is accurately shown, and the morphology of the radula, jaw and pseudobranch, as well as the reproductive system of three Mediterranean European locations studied are described. Conchological measurements were obtained with the help of the Image Kappa 5.1 software coupled to the high resolution binocular loupe, forming part of an ICAS system (Computer Image Analysis System). Both the study and the drawings of the reproductive system were made using the Leica M3Z stereomicroscope with clear coupled camera Wild 308700. For the study of both the ornamentation of the surface of the shells and the radula, the jaw and the male sexual pore photographs were made in the Scanning Electron Microscopy (SEM) Hitachi S–4100 with the digital program Sprit 1.8 (Bruker) and the Imagines application. All the material was coated with a layer of gold–paladium, for its exposure to the electron beam. The protoconch, teleoconch, radula and jaw studied with SEM correspond to specimens of the population of El Ejido (Almería). In the anatomical description of the different organs of the genitalia the use proximal denotes the part which is closer to the gonad, and distal the part which is closer to the gonopore.

*Shell growth in captivity*

30 specimens of *B. truncatus* were studied in captivity to determine their growth between March 2015 and June 2016. They were collected in December 2014 in El Ejido (Almería, Spain). The bulinids were transported under isothermal conditions to our laboratory. The possible natural infestation of schistosome of each of the collected specimens was checked individually, keeping them in Petri dishes with a small amount of mineral water. They were acclimated and maintained under controlled experimental conditions of 20ºC, 50% relative humidity and a photoperiod of 12 h light / 12 h dark in climatic chambers (model Heraeus–Vötsch VB–0714). The snails were placed in completely transparent plastic boxes inside which a silicone tube was inserted that releases air to keep the water as oxygenated as possible. At the bottom of the culture boxes was a substrate layer of calcareous stones submerged in mineral water. Their diet was based on lettuce (*Lactuca sativa*) for adults and *Oscillatoria formosa* cyanophycea algae for juveniles. From three specimens collected in El Ejido three generations were obtained in the laboratory. From each of these generations 30 specimens were selected. During the experiment, these specimens were kept individually in Petri dishes of 5cm in diameter with mineral water, fed with a small portion of lettuce, according to the size of the snail, and also with an aliquot of the *O. formosa* algae, in the case that they were recently hatched specimens. The length was monitored by biweekly scans of the 30 specimens in ventral and apical positions, with the help of the Image Kappa 5.1 previously cited. The morphometric measurements were obtained through the software Image Pro Plus 5.1, from the images previously generated.

**Results**

*Taxonomic data B. truncatus*

was collected from Egypt (*locus typicus*) and figured for the first time by Savigni (1817: pl. 2, fig. 27) without specifying the collection location, to which it does not assign any specific name. Later, Audouin (1827: fig. 29) comments on the species represented by Savigni and indicates that it is very similar to the *Physa truncata* of Férussac indicating conchological characters, although being an unpublished name corresponding to a *nomen nudum*, in consequence, the authorship of this taxon must be attributed to Audouin (1827) (Haas, 1935). Bouchet & Danrigal (1982: fig. 29) show a specimen of the Savigny collection, deposited in the Musée National d'Histoire Naturelle of Paris, which must be considered a lectotype of the species, corresponding to the other specimen deposited in that collection as a paralectotype. The type species of the genus *Bulinus* is *B. senegalensis* O.F. Müller, 1781 and its *locus typicus* Podor (Senegal). To systematic level Germain (1931) designates the family Bulinidae P. Fischer & Crosse, 1880 to group the genus *Bulinus* O.F. Müller, 1781 and genus *Indoplanorbis* Annandale & Prashad, 1921. Some authors had considered this valid (Larambergue, 1939; Baker, 1945) based on two characteristic features of the group and separate it from any other known division of the Planorbidae Rafinesque, 1815 based on the fluted or lobular pseudobranch (Fig. 21) and the shape of the penial complex in the genitalia. However, the pseudobranch in Planorbidae (comprising SF. Bulininae, now F. Bulinidae) is smooth and shaped like a leaf or square (Baker, 1945) (Fig. 22). Recently, Bouchet *et al*. (2017) based mainly on Morgan *et al.* (2002) and Aldrecht *et al*. (2007), point out that Bulinidae and Planorbidae are different families belonging to Superfamily Lymnaeoidea Rafinesque, 1815 and included in the Superorder Hygrophila Férussac, 1822.

In Europe two subspecies of *B. truncatus* are known: *B. t. contortus* (Michaud, 1829) from southern France and *B. t. rivularis* (Phillipi, 1836) of Sicily (Italy). Falkner *et al*. (2002) consider *Physa contorta* a subspecies of *Bulinus truncatus* while Boyer & Audibert (2007) consider *B. contortus* a junior synonym of *B. truncatus*, based on the fact that the shells of both taxa are indistinguishable and because the dispersion by birds has played an important role in the ubiquistic distribution of *B. truncatus*. This character is also applicable for *B. truncatus rivularis*, which would justify its presence in Sicily. Until now, only the population of El Ejido (Spain) (Figs 1–3) has been molecularly studied by sequencing *cox*1 (GenBank Acc. No. KR108924), confirming its specific assignment to *B. truncatus*, with 100% homology with other populations of this species available in GenBank. These first molecular data are the object of a more extensive study together with epidemiological studies (Martínez–Ortí *et al.*, 2015).

*External and internal anatomical data*

A regular dark brown in colour and with thin and elongated tentacles, shorter than in *Physa acuta* (Draparnaud, 1805), which also has a much lighter colour all over. It has the reproductive system, respiratory system (pneumostoma and pseudobranch) and excretory organs on the left side of the body. The displacement on the substrate is slow, unlike *P. acuta*, with which it is often confused, which is much faster. Planorbids are characterized by having a basic haploid chromosome number of 18 and the bulinid *B. truncatus* is considered polyploid (2n = 36) (Burch, 1960) and tetraploid (2n = 72), the latter associated with aphally (Brown, 1994). This tetraploid form is common among species of *Bulinus* which are the intermediate hosts of *S. haematobium* and cause urogenital schistosomiasis in humans (Burch, 1960; Davis, 1978; Dillon, 2004).

*Conchological characters (Figs 1–19, Table 1)*

The genus *Bulinus* presents about thirty species described, characterized by having a sinistral shell with a spire that is very variable in size and shape, a relatively high opening, evenly curved turns, and sometimes angular and but rarely warped (Brown, 1994; Martínez–Ortí *et al.,* 2015, 2019). The shell of *B. truncatus* is sinistral and has a very fragile ovoid globose shape, of pale amber colour, short spire, composed of 3–4 domed convex rounds, with deep sutures, and the latter being a little higher than half the shell and sculptured with thin radial ribs, although in general not very marked, thin, uneven and somewhat flexible. The first whorls of the shell of *B. truncatus* are ribbed. Blunt apex, oblique oval opening and slightly angled on the upper part, with the thin, subcontinuous peristome not reflected, without thickening, with the columellar margin usually narrow and more or less coiled and with narrow navel (Brown, 1994; Giusti *et al*., 1995; Martínez–Ortí *et al*., 2015, 2019) (Figs 1–19).

The maximum dimensions known of this species reach 13.0mm in height and 8.0mm in diameter (Germain, 1931; Nobre, 1941) (Table 1), although some authors point out that they can reach 20–23mm

(Malek & Cheng, 1974; Brown, 1994). The dimensions of the 10 specimens of the population of Villena (Alicante, Spain) (Figs 4–5) range between 5.2 and 7.1mm in height and between 3.5 and 4.4mm in diameter, while those in the population of El Ejido (Alicante, Spain) (Figs 1–3) range between 7.8 and 9.6mm in height and between 5.1 and 5.9 in diameter (Martínez–Ortí *et al.*, 2015, 2019). Of all the Spanish samples studied, the largest specimens reach 16.9mm in height and 11.6mm in diameter in two MNCN samples (MNCN–15.05/40640 and MNCN–15.05/26320). Nobre (1941) notes for Portuguese specimens 13.0mm in height and 8.0mm in diameter. Fraga de Azevedo *et al*. (1969) curiously consider that the Algarve specimens are larger than those of the north and centre of Portugal, due to temperature, obtaining 9.0mm maximum height and 5.5mm maximum diameter. The seven measured specimens of Corsica (Figs 12–13) range between 5.9mm and 8.45mm in height and between 4.1 and 5.0mm in width. The largest specimen presented the dimensions of $8.45 \times 5.0$mm and the smallest of $5.9 \times 4.0$mm. Ten specimens from Sardinia (Italy) (Fig. 11) have maximum measurements between 8.5 and 9.8mm in height and between 5.7 and 6.8mm in width, presenting the larger one the following dimensions $9.8 \times 5.9$mm and the smaller $8.5 \times 6.5$mm. Giusti *et al.* (1995) indicate dimensions for this species in the Maltese islands between 11.9–15.0mm in height and between 9.6–11.0mm in diameter. The maximum height can reach exceptionally 20mm (Giusti *et al*., 1995; Malek & Cheng, 1974). The features of the habitat (e.g. marshy bottom or not, abundance of food or scarcity, limestone or not, etc.) can influence both the size and shape of the shell (Schwetz, 1954).

According to the data obtained in Table 1 of 37 specimens the shell length (L) varies between 8.62±0.57mm in specimens from Almería and 16.12±0.65mm from Barcelona (for the 4 bigger specimens), and the height of aperture (HA)/ Length (L) relation of 33 specimens measured varies between 0.62 and 0.74. The height aperture (HA) of 33 specimens varies between 4.66±0.55 of Almería and 6.68±0.29mm of Sardinia. The data obtained of Fraga de Azebedo *et al*. (1969) was 0.57, although we do not know the number of specimens measured. These data indicate conchological variation among the different populations of this species, meaning one is unable to differentiate any of them by the shell. Although the population of Almería corresponds to the one with the smallest shell and the largest one to that of Barcelona populations. The protoconch, as in the genus *Bulinus* and the majority of planorbids, presents a punctiform microsculpture, whose spots appear arranged in spiral rows (Figs 14–15, 17–19) (Germain, 1931; Walter, 1962; Martínez–Ortí *et al*., 2015).

*Shell growth in captivity (Fig. 20)*

The growth in length of the shell has been followed in captive *B. truncatus* specimens from El Ejido (Almería, Spain). It has been shown that this growth is logistic type, since it increases rapidly from hatching and as the population approaches the carrying capacity of the environment, the growth rate becomes slower until it finally stabilizes (Molles, 2006). The shell size grows rapidly during the first ninety days, and during the following ninety days, the growth decreases and then from 150 to 180 days it increases scarcely. From September, the length of the shell continues to increase, although more slowly than at the beginning. After one year since hatching the growth is fully stabilized, giving rise to a linear growth during the following ninety days. Growth stabilization may be due, in part, to environmental resistance (Molles, 2006), which in our case includes abiotic factors such as temperature (20ºC), light (photoperiod 12 h light / 12 h dark), pH (close to 7), humidity (50%), properties of the mineral water in which they are kept and the type of nutrients that are supplied. A way of recognizing at the conchological level from the juvenile phase to the adult, whose growth is stabilized, is observing the ridge that appears in the peristome along the parietal and columellar edge of the opening (Figs 20A–C). In the most juvenile phase, 30 days after hatching, no thickening is observed in these areas of the opening (Fig. 20A). During the growth of the shell, these two zones thicken and leave a mark in the parietal and columellar areas of the opening (Fig.

20B, 240 days), and reaches a very clear edge or callus after the year of growth (Fig. 20C, 450 days), since the shell will swell more and more but without growing.

*Respiratory system (Figs 21–22)*

It is pulmonary and also consists of a pseudobranch on the left side of the body, that corresponds to an accessory gill that presents imbricate gill plates and is capable of capturing dissolved oxygen in the water while the animal is submerged but cannot obtain free oxygen. Traditionally, the pseudobranch is a taxonomic character having been used, together with the shape of the teeth of the radula and others, to differentiate the bulinids and planorbids (Germain, 1931; Baker, 1945). The rectum is placed above the pseudobranch and the anus opens between the pseudobranch and the pneumostoma, as in all planorbids and bulinids. Germain (1931) and Baker (1945) note that the shape of the pseudobranch is very characteristic in the bulinids, lobed and with folds, while in the planorbids it is compact, very prominent and not folded, smooth with variable shape, of leaf or square (Germain, 1931: p. 516; Baker, 1945: p. 50). In the two pseudobranchs observed in our specimens of *B. truncatus* from El Ejido, between 10 and 12 imbricate folds have been found (Fig. 21), being morphologically similar to those depicted by Hubendick (1978: fig. 88) for *Bulinus* sp. and by Stiglingh *el al*. (1962: fig. 19) and for *B. africanus* (Krauss, 1848). In *Helisoma duryi* (Wetherby, 1879) we have observed the two mentioned types. The only measured pseudobranch in the present work is of the square type (see Fig. 22), short and wide, and about 1.5mm long and 1.0mm high in size.

*Reproductive system (Figs 23–27)*

Hygrophilan snails are hermaphrodites, with spermatogenesis prior to oogenesis for at least a few weeks, not mating together at the same time. At least initially, in most cases, one snail must serve as the male and the other as the female. They have organs of both sexes, mostly separated, although other sexual organs are common (Baker, 1945). The reproductive system presents a general scheme of a diaulic and ditremated reproductive system, with the presence of two genital pores, one male and the other female (Fig. 26). However, like all freshwater pulmonates, they seem to be able to reproduce successfully by self–fertilization (Dillon, 1994).

In *B. truncatus* it is quite frequent to find aphallic specimens, and therefore monotremated, showing a scar on the surface of the skin produced by the closure of the male genital pore (Larambergue, 1939: fig. 20) (Fig. 28). The aphallic condition is also extremely common in the exanimate populations as well as the population from Sardinia (Italy), and the automixis could be the prevailing mode of reproduction (Wright, 1980; Lecis *et al*., 1984). Lecis *et al*. (1984) indicating that in populations of San Teodoro, Río Vignola in Posada (Sardinia), only aphallic specimens were found in 360 examined specimens. Fraga de Azevedo *et al*. (1969) found only aphallic individuals in Portugal, both in northern and southern populations. Larambegue (1939) studied two populations in Corsica (France), and although in Porto–Vecchio found both types in a euphallic/aphallic ratio 1:9.5 in Ajaccio they were discovered in 1:0.70. In Greece, this author only finds aphallic individuals in Heraklion, while in Kalamata both euphallics and aphallics individuals in ratio 1:5. The morphology of the aphallic specimens shows a reduction of the ducts and male organs, with no penial complex and with the vas deferens leading to the vagina from the prostatic gland (Larambergue, 1939) (Figs 23–25). Depending on the degree of reduction, the prostatic gland can also become rudimentary and the canal that connects it to the hermaphroditic canal corresponds to the vestige of the spermoviduct. Externally the aphally is translated in the absence of masculine pore, although a small whitish scar can be observed (Fig. 28).

The reproductive system of 20 specimens from El Ejido (Spain) and seven from Porto–Vecchio in Corsica (France) were examined and all found to be aphallics (Figs 23–25).

However, of 32 specimens of the Sardinia population examined, 7 euphallic and 25 aphallic specimens with a ratio 1:3.6 were found (Figs 26–27). Both reproductive morphologies, of euphallic and aphallic specimens, have been scarcely illustrated up to now, with only two papers reporting where euphallic specimens of this species appear (Larambergue, 1939; Malek & Cheng, 1974). The morphology of an euphallic specimen is characterized by the presence of the preputium, masculine organ more distal to the arrival of the phallotheca, which houses the retracted penis (Fig. 27), and goes outside through the proximal preputium that leads to the outside through the male genital pore. From the prostatic gland, placed on the uterus, the vas deferens arrives to the phallotheca (Fig. 26). All these male organs do not occur in an aphallic individual, in which the vas deferens coming from the reduced prostatic gland empties directly into the vagina (Figs 23–25).

*Jaw and radula (Figs 29–37)*
The jaws and the radula examined come from El Ejido (Almería, Spain). *B. truncatus* feeds on algae, ostracods, decaying organic matter and macrophytes, which are found on the firm mud in the bed of the water body (Dillon, 2004). The jaw is of polyplacognata type, formed by diverse fused plates that confer striations in its outer face and a little bent down by the ends and equal for the species of the genus (Germain, 1931; Hubendick, 1978: fig. 102) (Figs 29–30). No big differences are observed with those of other planorbids of the subfamily Helisomatinae (Baker, 1945). They are covered with thick cuticle to facilitate the cutting of leaves, filaments and other large particles (Dillon, 2004).

The radula has even been used to distinguish a group of species of the genus *Bulinus* (Schutte, 2009). The number of rows found in the three examined European radula of *B. truncatus* is variable, depending on the new rows that appear from the odontophore. Two of them have up to 117 rows and the other up to 120 rows (Fig. 31). Stiglingh *et al*. (1962) show that *B. tropicus* reaches 123 rows. The three radula of the Spanish population examined have the same radular formula: 29L + C + 29L. Burch & Kye–Heon (1984) indicate for *B. truncatus* specimens from Sudan a radular formula of 31L + C + 31L. The central tooth is bicuspid, with two cusps similar in size and with denticulate borders, arranged in the middle line of the radula and much shorter and narrower than the rest (Baker, 1945; Burch & Kye–Heon, 1984) (Figs 32–33). Burch & Kye–Heon (1984) indicate the presence in some occasions of a small central cusp to which they call "interstitial cusp", located between the two largest cusps, although it has not been observed in the radula examined by us. The lateral teeth closest to the central tooth are basically tricuspid, with a central triangular mesocone with arrowhead shape much larger than the two lateral cusps (endoconus and ectoconus) (Figs 32–33). As we progress towards the edges of the radula, the endoconus lengthens, occupying the inner and central margins of the tooth, while both the mesocone and the ectocone shortens and divides forming small cusps and occupying the lateral, external margin of the tooth (Figs 32–33). This dental transition to the margin is gradual, not observing sudden changes in its general form (Figs 34–35) (Burch & Kye– Heon, 1984). The marginal teeth, that occupy the outermost margin, are much narrower, elongated and minutely denticulate (Figs 36–37).

*Ecological data B. truncatus*
is an ubiquistic species with great capacity of self–fertilization that lives in coastal environments, whether they are lagoons, ponds, upwelling or river mouths where there is hardly any current or low flow–rate. The living Spanish population of *B. truncatus* studied comes from the pond of El Ejido (Almería). The snails have been found living on the surface of aquatic plants, with submerged or floating leaves of the phanerogam *Stuckenia pectinatus* (Linnaeus, 1753) and the green alga *Chara vulgaris* (Linnaeus, 1753) and on the mud of the bottom of the pond not far from the shore (Martínez–Ortí *et al*., 2015). Thomas (1995) indicates that *B. truncatus* lives on *Stuckenia* spp., *Ceratophyllum* spp. and *Lemna* spp. Like other intermediate fresh water snails hosts of *Schistosoma* it is able to tolerate desiccation in shaded temporary water bodies that gradually dried up (Chu *et al*., 1967; Brown, 1994; Van Aardt *et al*., 2007). The Almería pond considerably reduces its extension due to the low rainfall and high evaporation that it undergoes throughout the year, especially in the summer season, in which time *B. truncatus* can survive being buried in the mud (Ghandour, 1987; Van Aardt *et al*., 2007).

*B. truncatus* is a freshwater hermaphroditic snail with mean selfing rates exceeding 80% in natural populations. Therefore, when an individual of *B. truncatus* finds a new suitable ecosystem, the population grows rapidly and continuously, and allows it to expand its geographical area of occupation, through self–fertilization capacity, even becoming a pest, and its eradication is practically impossible. These isolated snails in the natural environment that grow more quickly can produce a much larger number of clusters, thus the population density increases and colonies originate, most likely as an effect of the absence of copulation (Bayomy & Joosse, 1987). The population of Almería was visited twice, once in 6$^{th}$ December 2014 when the population was abundant (approx. 50 specimens/m$^2$), with visible clusters and a large amount of water while in 8$^{th}$ July 2015, the water body was quite small and the population much smaller (approx. 6 specimens/m$^2$), with scarce clusters.

In nature, *B. truncatus* eggs require temperatures between 12.5ºC and 35ºC to hatch (McCrees h & Booth, 2013). Under laboratory conditions the isolated individuals start to lay eggs when they are approximately 5mm long (e.g. 35 days of age at 25ºC) (Doums *et al*., 1998). In our specimens the average value reached 5.31mm (n = 30). The physical–chemical parameters of water are important for the life of molluscs. A pH lower than 5.2 and high salinity inhibit the development of schistosomiasis vector snails (Okland, 1990; Thomas, 1995). In the south of Tunisia *B. truncatus* lives between 1220 and 2240 µS at 18ºC (Brown, 1994) and *B. obtusispira* less than 600 µS. The parameters obtained from the water in El Ejido (Almería) in December 2014 were water temperatures of 18ºC (11 h, with 16.0ºC) pH of 7.1 and a conductivity of 1250 µS, while in July 2015, with a large reduction of the pond extension, the data were as follows: dissolved oxygen: 2.5 mg/l (32%), temperature of water 34.6ºC (13:45 h, with 29ºC temperature) and a conductivity of 550 µS.

It is of interest to point out that the longevity of the planorbids in the wild is not well understood, and they can probably live between four and five years (Baker, 1945). Some species in captivity like *Helisoma trivolvis* (Say, 1917) lives up to 16 months, *H. duryi* one year and *Planorbarius corneus* (Linnaeus, 1758) up to two years. In our laboratory with water temperature at 20ºC, *B. truncatus* from Almería (Spain), reach almost two years of life, and between a month and a month and a half reach sexual maturity.

During our visits to the El Ejido pond (Almería), between July 2015 and the end of 2019, we identified some bird species of interest: the Eurasian coot (*Fulica atra* Linnaeus, 1758), the little egret [*Egretta garzetta* (Linnaeus, 1766)], the grey heron (*Ardea cinerea* Linnaeus, 1758) and the black–winged stilt [*Himantopus himantopus* (Linnaeus, 1758)].

These bird species are of interest because they can also be implicated as passive transporters in the passage from Africa to Southern Europe, be they lay, juveniles and even adults of *B. truncatus*, on their wings, feathers or legs (Martínez–Ortí *et al*., 2019).

Besides, the pond is often visited daily by a flock of about 300 sheep and some goats. The pond that we have recorded in these last two years at least has not never dried up. The water comes from the rains, from the surpluses of the greenhouses and from a broken drain pipe from the nearby hospital.

**Discussion**

Because of the conchological characteristics of *B. truncatus*, specimens can be confused with those of the physid *Physa acuta*, an exotic species widely distributed throughout Europe, with which it can coexist. Walter (1962) besides showing the punctiform ornamentation of *B. truncatus* of Sardinia and Egypt also shows that of *B. forskalii* (Ehrenberg, 1831) and *B. globosus* (Morelet, 1866) and even that of the genus *Indoplanorbis*. Walter (1962) points out the great importance of this conchological character, of great help to differentiate the hygrophilan snails with high shells, that transmit schistosomiasis from those that do not, as those species belonging to the family Physidae Fitzinger, 1833, whose protoconch is smooth (Martínez–Ortí *et al*., 2015).

According to the reproductive system the aphallic condition is predominant in *B. truncatus* and its acquisition of aphally is still unclear (Leonard & Córdoba–Aguilar, 2010). The production of aphallic specimens in *B. truncatus* involves both genetic factors, and other environmental factors such as higher temperatures, which increase the appearance of aphallic specimens, as also happens in reptiles, or exposure to light in slugs (Nicklas & Hoffmann, 1981; Doums *et al*., 1996; Dillon, 2004). An aphallic specimen can only receive sperm and auto-f ertilize, acting then as a female, when copulation occurs with a euphallic specimen (Pokryszko, 1987). After the first laying, the aphallic specimens of Posada (Sardinia) retain, in both the gonad and bursa copulatrix, the mature sperm (Lecis *et al*., 1984). The euphallic specimens seem to have mechanisms that favour fertilization by allosperm (donated by a partner) over autosperm. Autosperm, produced endogenously, is generally transported through the male duct into the vas deferens, along a prostatic gland, to the penis (Dillon, 2004). *Bulinus* can store sperm over seven weeks of starvation, eight weeks of low temperature, or four weeks of desiccation (Rudolph & Bailey, 1985). However, the euphallic forms are also known in *B. truncatus* (Larambergue, 1939; Malek & Cheng, 1974; Jarne *et al*., 1992). In Bulininae, the male reproductive system of euphallic specimens is very characteristic, and the penial complex next to the prostatic gland has great taxonomic importance (Baker, 1945) (Fig. 26).

Infection of molluscs by digenean trematode parasites typically results in the repression of reproduction, so–called parasitic castration. The aphally can be an adaptation to reduce the consequences of parasitism of flukes (Schrag & Rollinson, 1994). Parasitic castration results in the penis being greatly reduced, as is the whole reproductive system, and sometimes complete destruction of the gonads. This is known to occur by altering the expression of a range of host neuropeptide genes (Adiyodi & Adiyodi, 1994; Rice *et al*., 2006). The size and prominence of the accessory sex organs (penis and vas deferens in male) become abnormal after the invasion of the gonads by parasites. It seems to indicate that the size of these sex organs depends on gonadal activity (Reeves, 1936; Leonard & Córdoba– Aguilar, 2010). Rotchschild (1938) found in *Hydrobia ulvae* (Pennant, 1777) parasitized specimens with abnormal penises, with sizes more reduced (hemiphally), and absence total of penis due to the total destruction of the testicle (aphally). However, the female organs and ducts are perfectly normal and allow the reproduction through self-f ertilization, a unique way of reproduction for such specimens. The aphallic specimens are more fertile and lay more eggs than the euphallics, important for the colonization of new places. The aphally is transmitted preferably to its descendants, although they can also produce euphallic descendants (Larambergue, 1939; Jarne *et al*., 1992; own dates). Although initially it was thought that these individuals can invest more energy in reproduction, avoiding it in the formation and maintenance of the male organs (Jarne *et al*., 1992), later Ostrowski *et al.* (2003) suggest that the cost of the construction and maintenance of the phallus must be very low in this species and suggest investigating the cost associated with the use of the penis (e.g. search of a mate, courtship and copulatory behaviours, sperm production) to explain the maintenance of high frequencies of aphallic individuals in natural populations.

**AcknowledgMents**

To Dr. Francesc Uribe, curator of molluscs, and Miguel Prieto, data manager, of the "Museu de Ciències Naturals" of Barcelona, and to Dr. Rafael Araujo, curator of molluscs of the "Museo Nacional de Ciencias Naturales" of Madrid, who we remember very emotionally after his recent passing, for allowing us to review the samples deposited in their respective collections. Also, Dr. Elisabetta Pintore from the Università degli Studi di Sassari of Sardinia (Italy) for help with providing us with material of *B. truncatus*. Finally, to the Section of Electron Microscopy of the SCSIE of the University of Valencia for assistance with photography using the MEB Hitachi S–4100.

This study was financed by Health Research Project No. PI16/00520, Subprograma Estatal de Generación de Conocimiento de la Acción Estratégica en Salud (AES) y Fondos FEDER, Plan Estatal de Investigación Científica y Técnica y de Innovación, ISCIII–MINECO, Madrid, Spain; by the Red de Investigación de Centros de Enfermedades Tropicales–RICET (Project No. RD16/0027/0023 of the PN de I + D + I, ISCIII– Subdirección General de Redes y Centros de Investigación Cooperativa RETICS), Ministry of Health and Consumption, Madrid; by PROMETEO Programs, "Programa de Ayudas para Grupos de Investigación de Excelencia", Generalitat Valenciana, PROMETEO/2016/099 and PROMETEO/2021/004, Valencia, Spain. Also funded by CIBER de Enfermedades Infecciosas (CB21/13/00056), ISCIII, Ministry of Science and Education, Madrid, Spain.

**Table 1** Shell size of *Bulinus truncatus* populations in different countries of the western Mediterranean frame. Values in mm. (–) data not available.

| | | Shell Length (L) | | Shell Width | | Height of Aperture (HA) | | HA/L | |
|---|---|---|---|---|---|---|---|---|---|
| Population | n | range | mean±SD | extreme values | mean ±SD | extreme values | mean ±SD | Relation | References |
| El Ejido (Spain) | 10 | 7.8–9.6 | 8.62±0,57 | 5.1–5.9 | 5.45±0.28 | 4.75–6.1 | 5.37±0.41 | 0.62 | present study |
| Villena (Spain) | 6 | 5.2–7.1 | 6.43±0,85 | 3.5–4.4 | 4.16±0.50 | 3.7–5.1 | 4.66±0.55 | 0.72 | present study |
| Corsica (France) | 7 | 5.9–8.5 | 7.0±0.9 | 4.1–5.0 | 4.52±0.5 | 4.45–5.6 | 4.8±0.45 | 0.68 | present study |
| Sardinia (Italy) | 10 | 8.5–9.8 | 8.96±0.43 | 5.7–6.8 | 6.18±0.36 | 6.3–7.3 | 6.68±0.29 | 0.74 | present study |
| Barcelona (Spain); MNCN samples | 4 | 15.5–16.9 | 16.12±0.65 | 10.6–11.6 | 11.07±0.55 | – | – | – | present study |
| Palma de Mallorca (Spain); MNCN-15.05/40630 | 1 | 14.0 | – | 9,2 | – | – | – | – | present study |
| North and Center Portugal | – | 13.0 | – | 8.0 | – | – | – | – | Nobre (1941) |
| Algarve (Portugal) | – | 9.0 | – | 5.5 | – | 5.2 | – | 0.57 | Fraga de Azevedo et al. (1969) |
| North Portugal | – | 6.7 | – | 4.6 | – | 4.2 | – | 0.62 | Fraga de Azevedo et al. (1969) |
| Malta island | – | 11.9–15.0 | – | 9.6–11.0 | – | – | – | – | Giusti et al. (1995) |

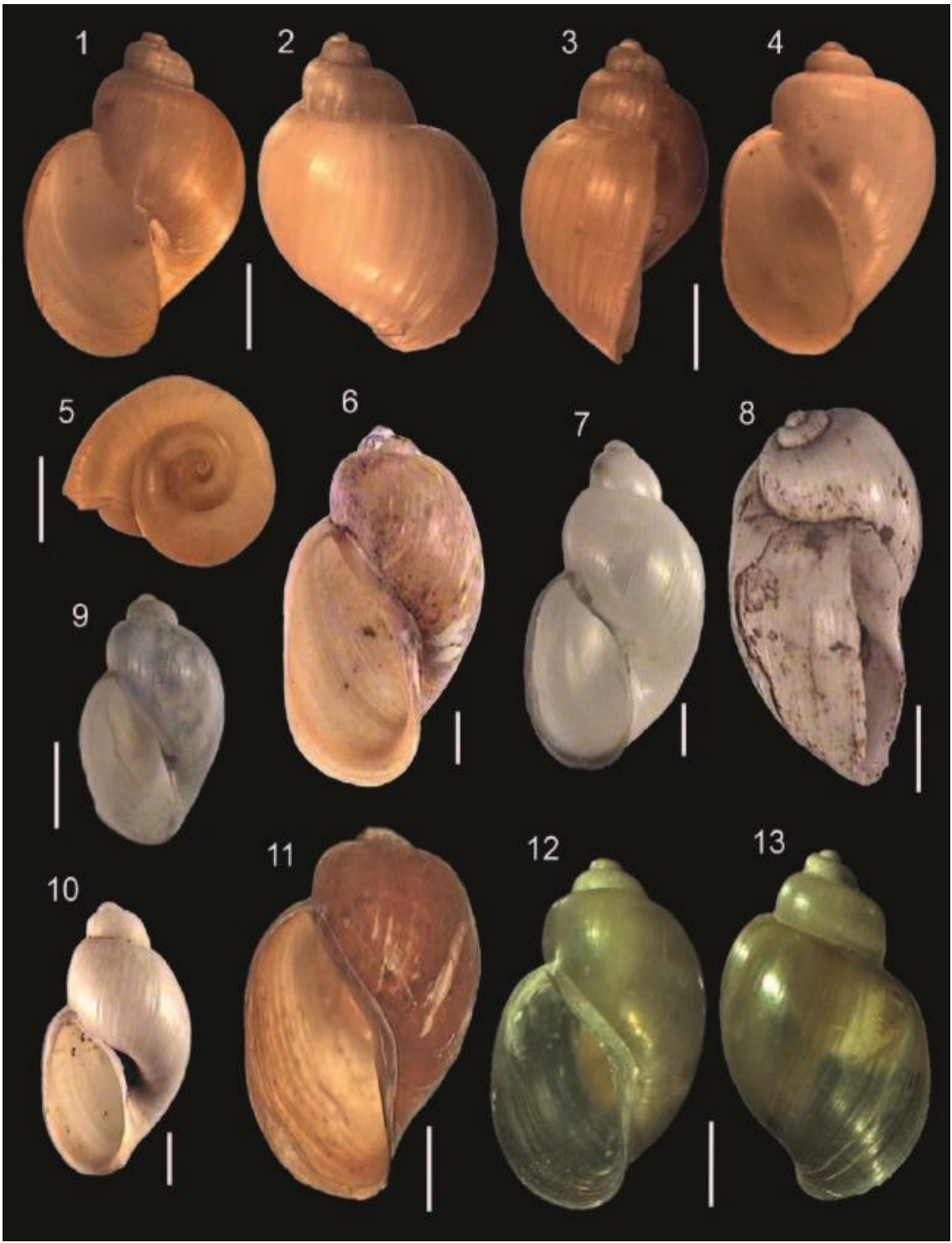

**Figures 1–13** Shells of *Bulinus truncatus* of Spain, Italy and France: **1–3** El Ejido (Almería, Spain) (MVHN–071214TP04) **4–5** Villena (Alicante, Spain) (MVHN–281214TY01) **6** Barranc de l'Algendar (Menorca, Balearic Islands, Spain) (MZB 80–3053) **7** Ampurias (Girona, Spain) (MZB 2009–0536) **8** San Antonio Abad, Ibiza (Balearic Islands, Spain) (MZB 84–1825) **9** Adra (Almería, Spain) (MZB 84–7124) **10** Font de Mestre Pere, Mallorca (Balearic Islands, Spain) (MZB 84–1822) **11** San Teodoro (Sardinia, Italy) (MVHN–031215MA06) **12–13** Porto–Vecchio, Corsica (France) (MVHN–240815AF01). Scales = 2mm.

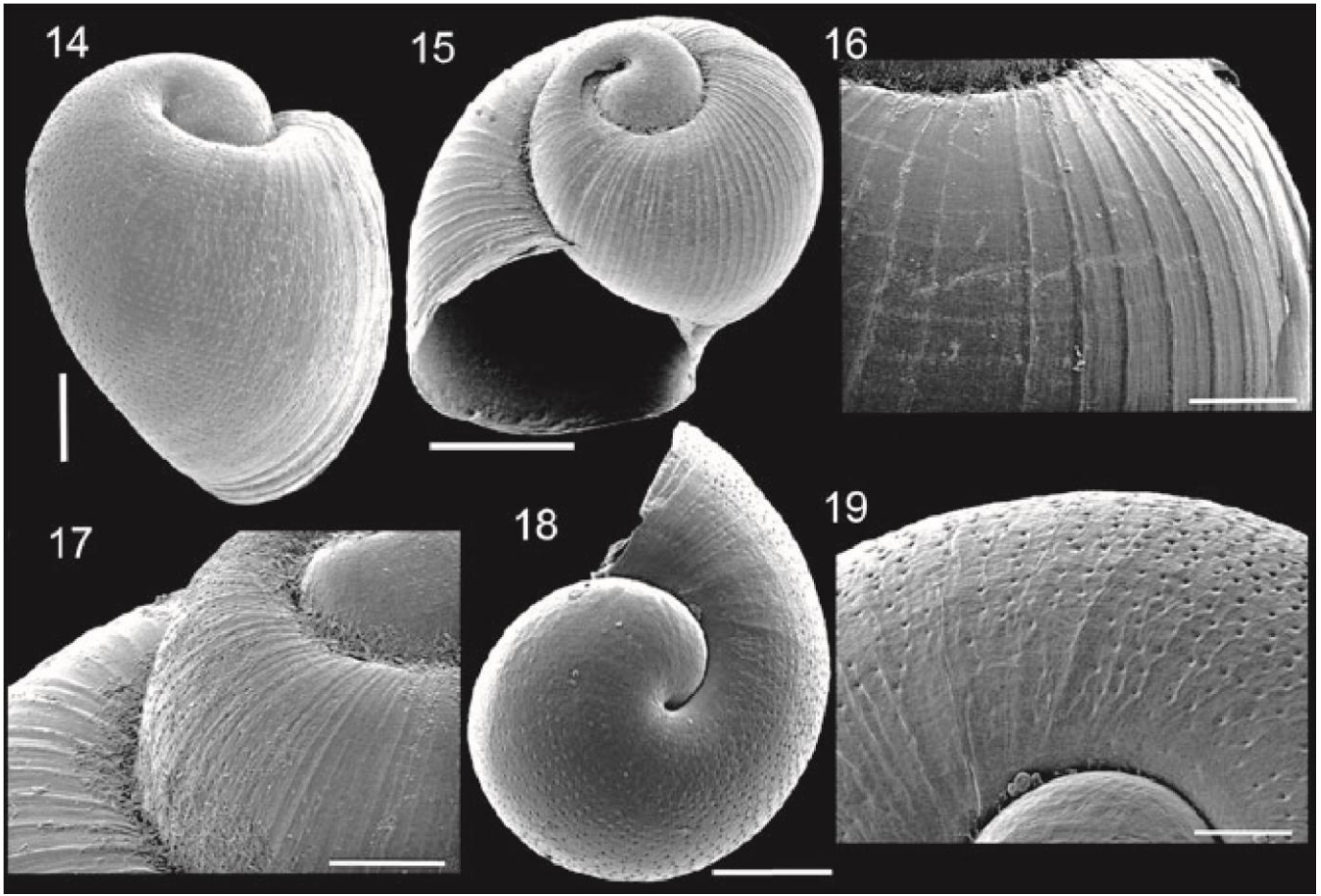

**Figures 14–19** Microsculpture of the protoconch and the first whorls of *B. truncatus* (El Ejido, Almería, Spain). Scales: 14,17,18, = 200μm; 15, = 600μm; 16, = 300μm; 19, = 70μm.

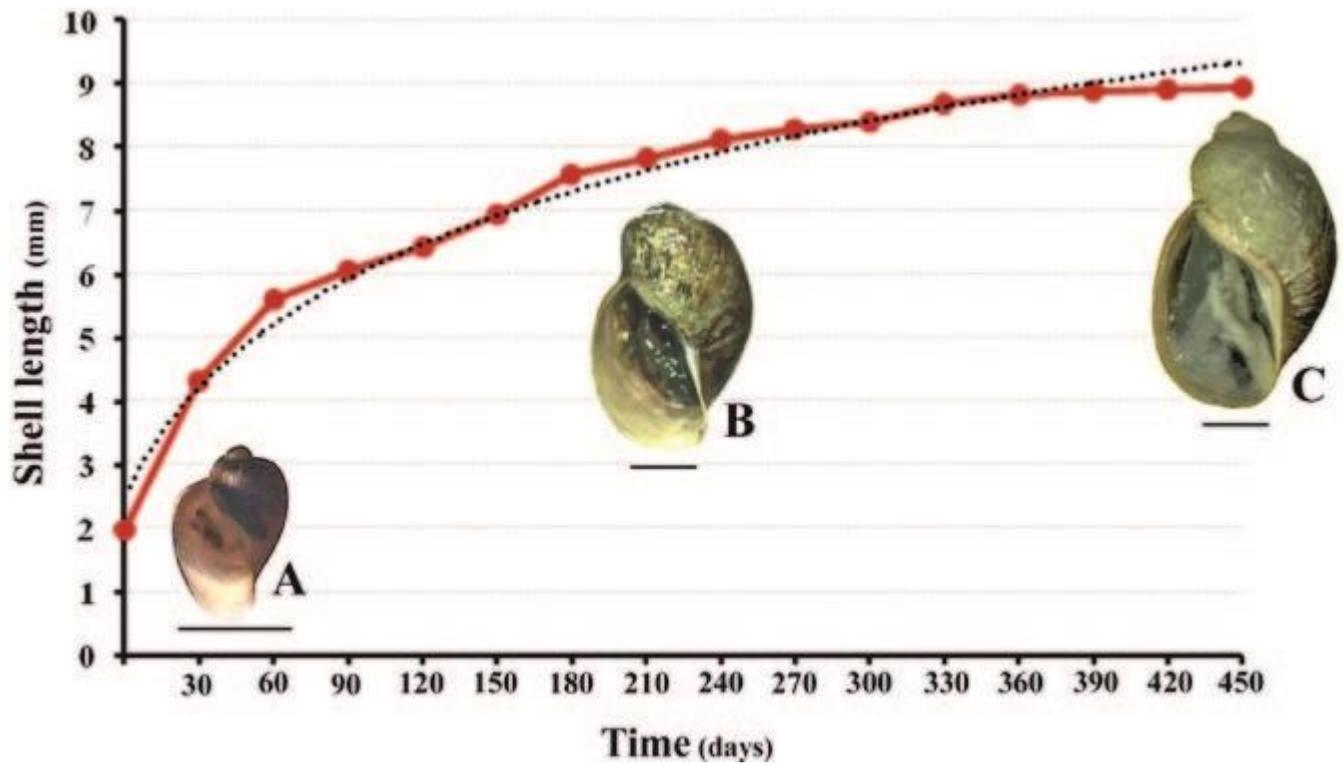

**Figure 20** Variation of the length of the shell (in mm) of *Bulinus truncatus* with respect to the time in days, from El Ejido (Almería, Spain). Scales = 2mm.

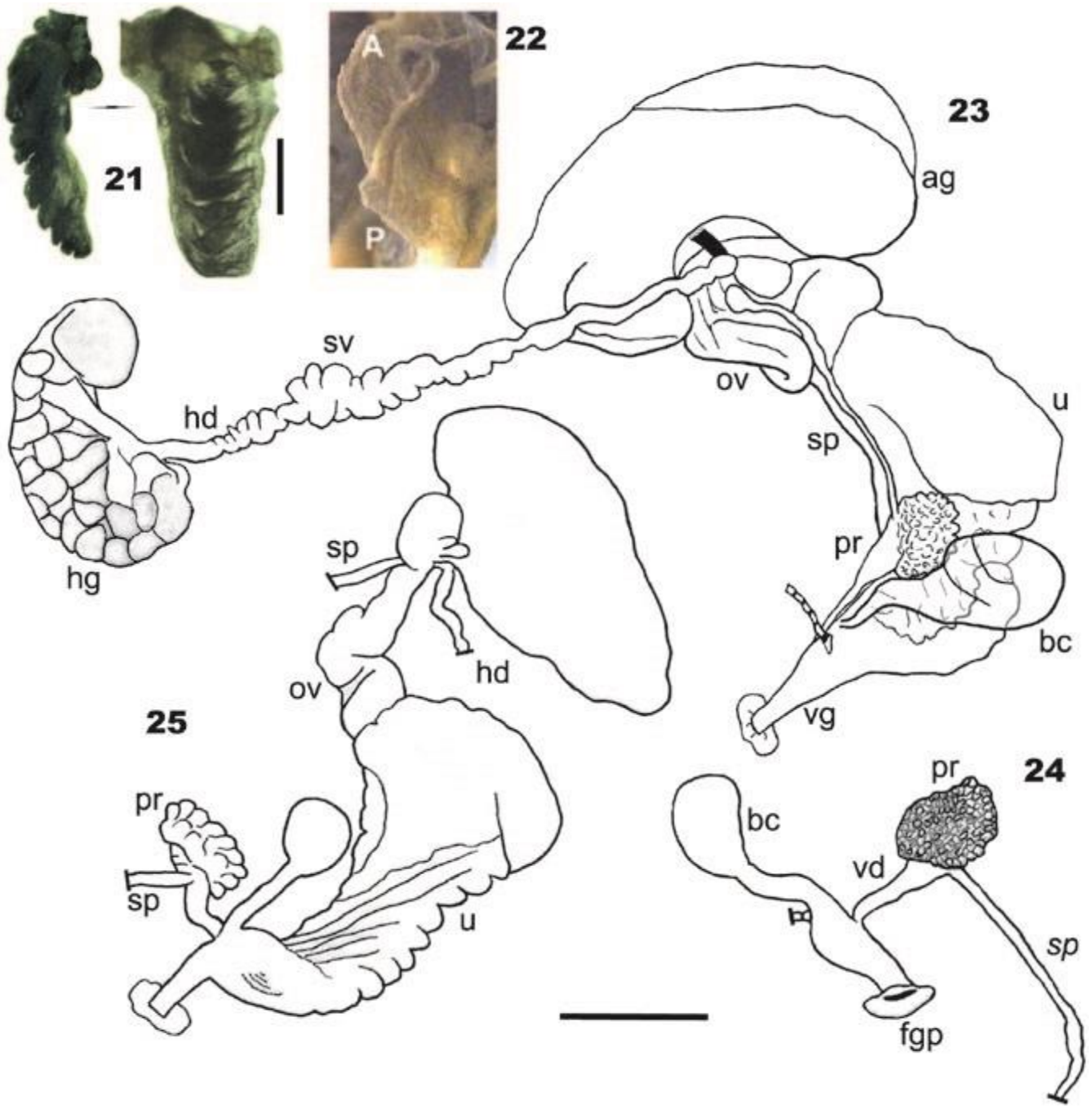

**Figures 21–25** Pseudobranch and aphallic reproductive system of *Bulinus truncatus* and the pseudobranch of *Helisoma duryi* **21** Pseudobranch of *B. truncatus*. El Ejido (Almería, Spain) **22** Pseudobranch of *H. duryi*. Ágora (Valencia, Spain) **23–24** Reproductive system of two specimens of *B. truncatus* of El Ejido (Almería, Spain) **25** Specimen from Porto–Vecchio. Corsica (France) (abbreviations: a = anus; ag = albumen gland; bc = bursa copulatrix; fgp = female genital pore; hd = hermaphroditic duct; hg = hermaphroditic gland; ov = oviduct; p = pseudobranch; pr = prostatic gland; sp = spermoviduct; sv = seminis vesicula; u = uterus; vd = vas deferens; vg = vagina). Scales = 1mm.

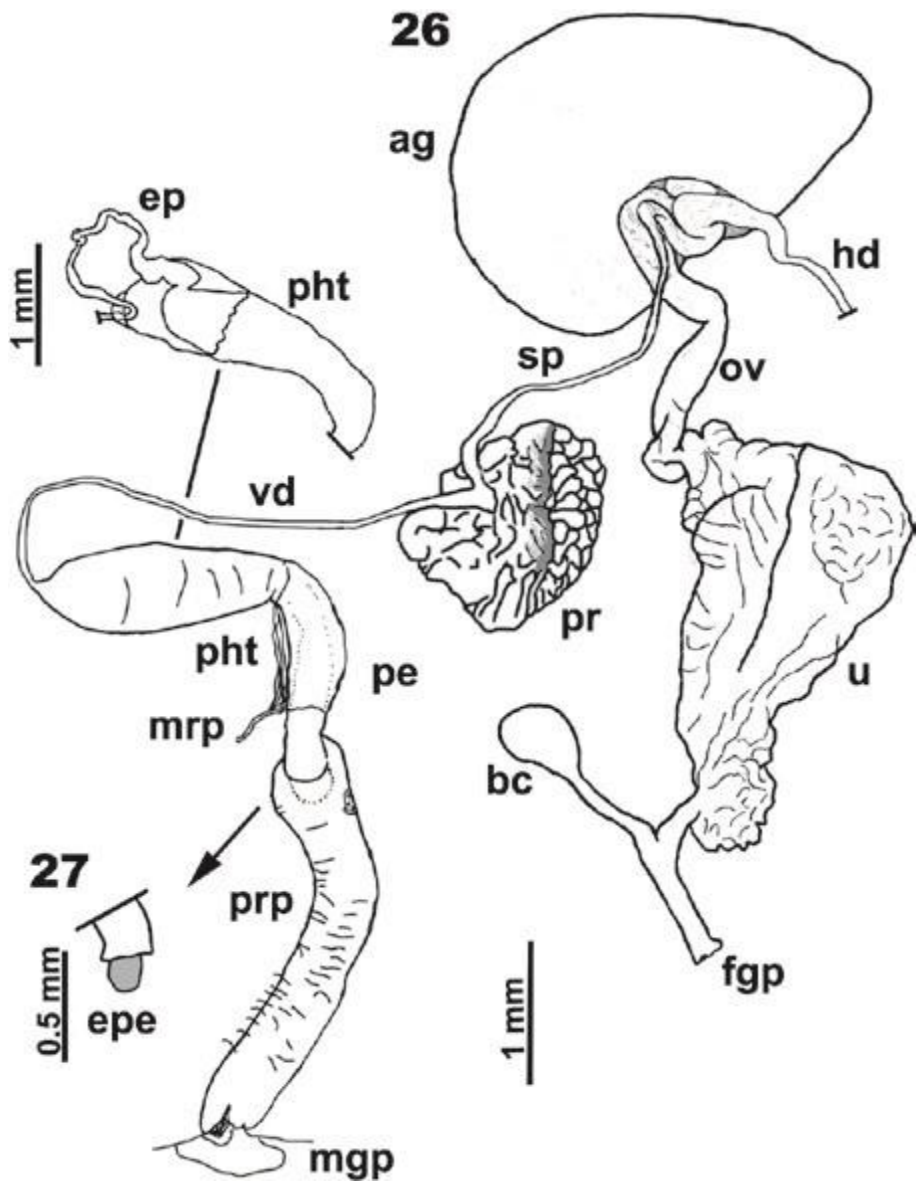

**Figures 26–27 26** Reproductive system of a one euphallic specimen of San Teodoro (Sardinia, Italy) **27** Everted penis (abbreviations: ag = albumen gland; bc = bursa copulatrix; epe = everted penis; fgp = female genital pore; hd = hermaphroditic duct; hg = hermaphroditic gland; mgp = male genital pore; mrp = penis retractor muscle: ov = oviduct; pe = penis; pht = phallotheca; pr = prostatic gland; prm = penis retractor muscle; prp = preputium; sp = spermoviduct; u = uterus; vd = vas deferens). Scales: 26, = 1mm; 27, = 0.5mm.

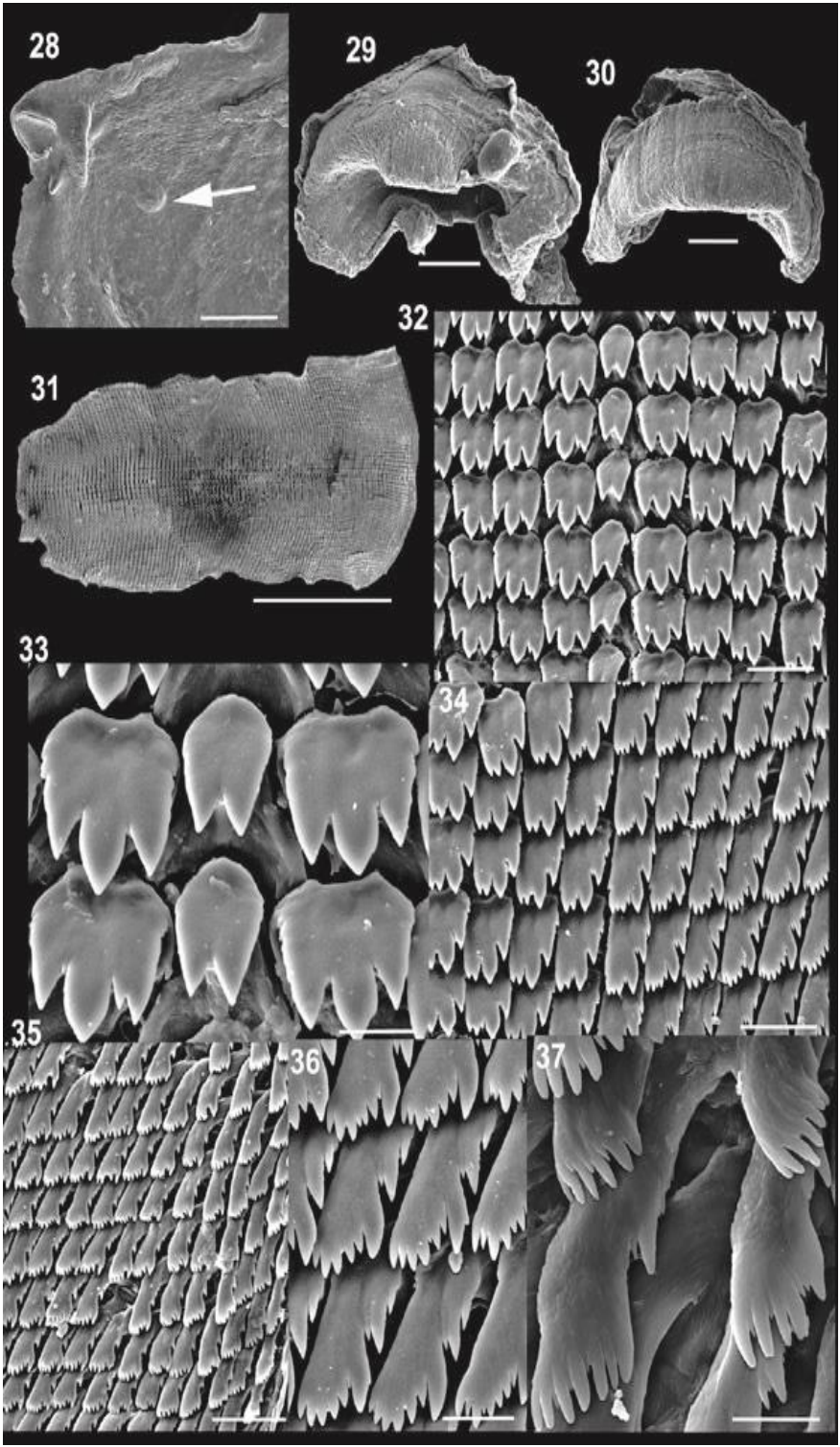

**Figures 28–37** Some morphological characters of *Bulinus truncatus* of El Ejido (Spain) **28** Closed male genital pore **29–30** Jaws **31** General view of radula **32–37** Radula details **32** Central and first lateral row **33** Central and first lateral teeth **34–35** Denticular transition of lateral teeth towards the margin of the radula **36–37** Detail of marginal teeth.
Scales: 28, = 300µm; 29,30, = 50µm; 31, = 500µm; 32,34, = 20µm; 33, = 9µm; 35, = 30µm; 36, = 10µm; 37, = 6µ.